\documentclass[prd,reprint,twoside,letterpaper,showpacs,amsmath,nofootinbib,superscriptaddress]{revtex4-1}

\usepackage[T1]{fontenc}
\usepackage[utf8]{inputenc}
\usepackage[brazil,english]{babel}
\usepackage{tensor}
\usepackage[usenames,dvipsnames]{color}
  \definecolor{darkblue}{RGB}{0,0,150}
\usepackage{amssymb} 
\allowdisplaybreaks[1]
\usepackage{graphicx}
\usepackage{nicefrac}
\usepackage{paralist}
\usepackage{fancyhdr}
\makeatletter
  \fancypagestyle{plain}{\fancyhead{}}
\makeatother
\usepackage[unicode,pdfusetitle]{hyperref}
\makeatletter
\hypersetup{
  pdfinfo = {
    Author = {Daniel C. Guariento, Flavio Mercati},
    Title = {\@title},
    Subject = {Self-gravitating systems; continuous media and classical fields in curved spacetime; exact solutions; Quantum gravity; classical black holes}
  },
  breaklinks = true,
  colorlinks = true,
  linkcolor = darkblue,
  urlcolor = darkblue,
  citecolor = darkblue
}
\makeatother

\newcommand{\ud}{\ensuremath{\mathrm{d}}}
\newcommand{\sfrac}[2]{\ensuremath{\textstyle{\frac{#1}{#2}}}}

\makeatletter
\providecommand*{\diff}%
{\@ifnextchar^{\DIfF}{\DIfF^{}}}
\def\DIfF^#1{%
  \mathop{\mathrm{\mathstrut d}}%
  \nolimits^{#1}\gobblespace}
\def\gobblespace{%
  \futurelet\diffarg\opspace}
\def\opspace{%
  \let\DiffSpace\!%
  \ifx\diffarg(%
  \let\DiffSpace\relax
  \else
  \ifx\diffarg[%
  \let\DiffSpace\relax
  \else
  \ifx\diffarg\{%
  \let\DiffSpace\relax
  \fi\fi\fi\DiffSpace}

\DeclareMathOperator{\diag}{diag}

\begin{document}

\title{Self-gravitating fluid solutions of Shape Dynamics}

\author{Daniel C.\ Guariento}
\email{dguariento@perimeterinstitute.ca}

\affiliation{Department of Physics \& Astronomy, University of Waterloo, Waterloo, ON, N2L 3G1, Canada}
\affiliation{Perimeter Institute for Theoretical Physics, 31 Caroline St.\ N., Waterloo, ON, N2L 2Y5, Canada}

\author{Flavio Mercati}
\email{fmercati@perimeterinstitute.ca}

\affiliation{Perimeter Institute for Theoretical Physics, 31 Caroline St.\ N., Waterloo, ON, N2L 2Y5, Canada}

\begin{abstract}

Shape Dynamics is a 3D conformally invariant theory of gravity which possesses a large set of solutions in common with General Relativity. When looked closely, these solutions are found to behave in surprising ways, so in order to probe the fitness of Shape Dynamics as a viable alternative to General Relativity one must find and understand increasingly more complex, less symmetrical exact solutions, on which to base perturbative studies and numerical analyses in order to compare them with data. Spherically symmetric exact solutions have been studied, but only in a static vacuum setup. In this work we construct a class of time-dependent exact solutions of Shape Dynamics from first principles, representing a central inhomogeneity in an evolving cosmological environment. By assuming only a perfect fluid source in a spherically symmetric geometry we show that this fully dynamic non-vacuum solution satisfies in all generality the Hamiltonian structure of Shape Dynamics. The simplest choice of solutions is shown to be a member of the McVittie family.

\end{abstract}


\maketitle

\section{Introduction}

Shape dynamics (SD) is a theory of gravity which is locally identical to the Arnowitt-Deser-Misner (ADM) formulation~\cite{adm-1962} of General Relativity (GR) in a constant-mean-extrinsic-curvature (CMC) foliation. In such a foliation the physical degrees of freedom of the gravitational field are 3-dimensional conformally invariant \cite{York:1971hw,*ChoquetBruhat-1973,*O'Murchadha:1973yra,Gomes:2010fh,*Gomes:2011zi,Mercati:2014ama}. SD consists in taking this symmetry as fundamental, and requiring that the 3-dimensional conformal geometry of each CMC slice be regular. This gives rise to solutions that may differ from those of GR, in which the fundamental requirement is that of regularity of the pseudo-Riemannian geometry of 4-dimensional spacetime.

The relational principles on which SD is based~\cite{Mercati:2014ama} only require that its solutions be generated by a Hamiltonian that is invariant under 3D diffeomorphisms and conformal transformations of the spatial metric. This allows for a large class of Hamiltonians, generating inequivalent dynamics, but among which we have to select one based on consistency with observations~\cite{Mercati:2014ama,Gomes:2010fh,Gomes:2011zi}. The value of this Hamiltonian at each point of the phase space of ADM variables can be determined only by solving a quasilinear elliptic differential equation, whose result depends nonlocally on the ADM variables. This equation obviously cannot be solved exactly everywhere in phase space, and therefore, for practical applications, we need to compute it as a perturbative expansion. But any perturbative approach needs an exact solution to expand around, and for this reason exact solutions of SD are an invaluable starting point for any specific application of the theory. The simplest class of solution are the homogeneous ones, which have been studied in Refs.~\cite{Mercati:2014ama,UpcomingTimFlavioDavid}. Giving up homogeneity, we can assume spherical symmetry in order to make the Hamiltonian calculable, but in that case we also need to add some matter source, to have nontrivial shape degrees of freedom (any spherically symmetric manifold is conformally flat, so its shape configuration space is just a point). The simplest thing to do is to add a pair of thin shells of dust to a spherical universe~\cite{Mercati:2014ama,Gomes:2015ila}, a procedure which leads to a very nontrivial system. In this paper we are interested in coupling SD to a less singular kind of matter source: a spherically-symmetric perfect fluid. By not assuming anything on the relation between density and pressure of this fluid, we will be able to solve all the constraints of SD and get a class of exact solutions, which will be a very valuable starting point for further perturbative analyses. Furthermore, by borrowing a familiar assumption in GR, namely that the Weyl part of the Hawking--Hayward quasilocal mass takes the form that one expects in the presence of a (cosmological) black hole, or central mass, we are able to derive one particular exact solution which is known in GR as the McVittie metric. Such a metric describes a cosmological black hole in the presence of a fluid, and it is interesting to observe that is is a solution of SD too. We believe that the simplicity of the McVittie metric will make it a very interesting playground for SD, in particular for the understanding of its physical predictions regarding black holes. The McVittie metric, in fact, is a non-static cosmological solution of SD which has a central concentration of mass which we can identify with a black hole. The other candidate black hole solutions of SD that have been found so far~\cite{Gomes:2013fca,Gomes:2013bbl} are static, eternal and asymptotically flat, which makes them significantly less interesting from the physical point of view.

For practical purposes, instead of working directly with the SD dynamical system, it is convenient to simply work in the conformal gauge in which SD is equivalent to GR in a CMC foliation. We will therefore be studying solutions of Arnowitt-Deser-Misner gravity in CMC gauge. When such solutions exist, they are both solutions of GR and SD. However there are situations in which such solutions do not correspond to a well-defined solution of Einstein's equations, in particular at the Big-Bang singularity \cite{UpcomingTimFlavioDavid}. However, by looking at the conformally-invariant degrees of freedom, one can check whether, as solutions of SD, they still make sense and can be continued past this breakdown point. The strategy is to work with ADM gravity in CMC gauge as long as it is possible, and then focus on the shape degrees of freedom when the solutions evolve into something that cannot be described in GR.

In this work we derive a class of exact solutions of SD by solving the ADM equations under the following assumptions: 
\begin{inparaenum}[(i)]
\item spherical symmetry; \label{cond-sphsym}
\item comoving perfect fluid source; \label{cond-perf}
\item asymptotically FLRW behavior; and\label{cond-flrw}
\item a singularity at the center. \label{sing-center}
\end{inparaenum}
Conditions \ref{cond-sphsym} and \ref{cond-perf} define the Kustaanheimo-Qvist class of solutions, of which there are many physically interesting subcases \cite{kustaanheimo-1947}, such as Wyman \cite{wyman-1946}, FLRW, McVittie and Schwarzschild-de~Sitter \cite{stephani-exact}. Conditions \ref{cond-flrw} and \ref{sing-center} represent fixing properties of the poles of the spherically symmetric manifold. In particular, condition \ref{cond-flrw} requires that the metric be regular at one of the poles, and condition \ref{sing-center} requires it to be singular at the opposite pole.

In Sec.\ \ref{sec:eqmov} we use the constraints and equations of motion of Shape Dynamics to write the spherically symmetric ansatz in its specific form. Using conditions \ref{cond-sphsym} to \ref{cond-perf} we arrive at the generic expression for the Kustaanheimo-Qvist class of spherically symmetric geometries. In Sec.\ \ref{sec:ms-mass} we use the 3-dimensional expression of the Hawking--Hayward quasilocal mass to apply conditions \ref{cond-flrw} and \ref{sing-center}, and we show in Sec.\ \ref{sec:lapsefix} that the solution satisfies all the requirements of a solution of Shape Dynamics. We present our conclusions and discuss further developments in Sec.\ \ref{sec:conclusions}. Latin indices run from 1 to 3 and in our units $16 \pi G = 1$.

\section{Statement of the problem}\label{sec:eqmov}

The gravitational action of a system filled with a continuous fluid with an arbitrary energy momentum tensor is given by
\begin{equation}
  S = S_{\text{EH}} + S_\text{M} \,,
\end{equation}
where $S_\text{EH}$ is the Einstein-Hilbert action given by
\begin{equation}
  \begin{split}
    S_{\text{EH}} =&\, \int \diff t \int \diff^3 x \left\{ \pi^{a b} \dot{\gamma}_{a b} \vphantom{\frac{1}{\sqrt{\gamma}}} \right.\\
    &- N \left[ \frac{1}{\sqrt{\gamma}} \left( \pi^{a b} \pi_{a b} - \frac{1}{2} \pi^2 \right) - \sqrt{\gamma} \mathcal{R} \right] \\
    & \left. \vphantom{\frac{1}{\sqrt{\gamma}}} - 2 N_a \nabla_b \pi^{a b} \right\} \\
    =&\, \int \diff t \int \diff^3 x \left( \pi^{a b} \dot{\gamma}_{a b} - N \mathcal{H} -N_a \mathcal{H}^a \right) \,,
  \end{split}
\end{equation}
where $\mathcal{R}$ is the 3-Ricci scalar and we have the usual definitions for the ``superhamiltonian'' and ``supermomentum'', namely
\begin{align}
  \mathcal{H} \equiv&\, \frac{1}{\sqrt{\gamma}} \left( \pi^{a b} \pi_{a b} - \frac{1}{2} \pi^2 \right) - \sqrt{\gamma} \mathcal{R} \,,\\
  \mathcal{H}^a \equiv&\, 2 \nabla_b \pi^{a b} \,,
\end{align}
and the matter action $S_\text{M}$ is left unspecified for now. It may be possible to define the Hamiltonian of an arbitrary perfect fluid by defining it as a generic $k$-essence action. This procedure has been carried out for some particular cases \cite{Abramo:2005be}, but the general action problem will be addressed in a future work. We assume that it depends solely on the metric components and not on their associated momenta, so that we may define the components of the variation of $S_\text{M}$ with respect to the lapse, shift and metric as
\begin{align}
  \rho \equiv&\, -\frac{1}{N \sqrt{\gamma}} \frac{\delta S_\text{M}}{\delta N} \,,\\
  j_a \equiv&\, -\frac{1}{N \sqrt{\gamma}} \frac{\delta S_\text{M}}{\delta N^a} \,,\\
  S_{a b} \equiv&\, -\frac{1}{N \sqrt{\gamma}} \frac{\delta S_\text{M}}{\delta \gamma^{a b}} \,.
\label{eq:StressTerm}
\end{align}
The Hamiltonian constraint and momentum (or diffeomorphism) constraint are, respectively,
\begin{align}
  \mathcal{H} =&\, \sqrt{\gamma} \rho \,, \label{eq:ham-const}\\
  \mathcal{H}^a =&\, \sqrt{\gamma} j^a \,. \label{eq:mom-const}
\end{align}
The CMC condition reads
\begin{equation}\label{eq:cmc-cond}
  \gamma_{a b} \pi^{a b} - \sqrt{\gamma} \left< \pi \right> = 0 \,,
\end{equation}
where $\left< \pi \right>$ is the mean canonical momentum across the 3-surface, that is, $\left< \pi \right> \equiv \frac{\int \diff^3 x \gamma_{a b} \pi^{a b}}{\int \diff^3 x \sqrt{\gamma}}$. The evolution equations divide into a vacuum part, which is given by Hamilton's equations generated by the total Hamiltonian $\int \diff t \int \diff^3 x \left(N \mathcal{H} + N_a \mathcal{H}^a \right)$, and a matter contribution to $\dot{\pi}^{ab}$ given  by the term \eqref{eq:StressTerm}. The equations read \cite{adm-1962}
\begin{align}\label{eq:evolg}
  \dot{\gamma}_{a b} =&\, 2 \frac{N}{\sqrt{\gamma}} \left( \pi_{a b} - \frac{1}{2} \pi \gamma_{a b} \right) + 2 \nabla_{\left( a \right.} N_{\left. b \right)} \,,\\
  \begin{split}\label{evolp}
    \dot{\pi}^{a b} =&\, - N \sqrt{\gamma} \left( \mathcal{R}^{a b} - \frac{1}{2} \gamma^{a b} \mathcal{R} \right) \\
    & + \frac{N}{\sqrt{\gamma}} \left[ \frac{\gamma^{a b}}{2} \left( \pi^{c d} \pi_{c d} - \frac{1}{2} \pi^2 \right) \right.\\
    & \left. - 2 \left( \pi^{a c} \, \tensor{\pi}{_c ^b} - \frac{1}{2} \pi \, \pi^{a b} \right) \right] \\
    & + \sqrt{\gamma} \left( \nabla^a \nabla^b N - \gamma^{a b} \nabla^c \nabla_c N \right) \\
    & + \nabla_c \left( \pi^{a b} N^c \right) - 2 \pi^{c \left( a \right.} \nabla_c N^{\left. b \right)} + N \sqrt{\gamma} S^{a b} \,.
  \end{split}
\end{align}

In order for a solution of SD to admit a description as a spacetime solving Einstein's equation (\emph{i.e.} a solution of GR), a lapse function defining a local notion of proper time must be defined. Such a function can be determined by using the ADM equations of motion~\eqref{eq:evolg},~\eqref{evolp} to calculate the time derivative of the CMC condition~\eqref{eq:cmc-cond}:
\begin{equation}\label{eq:LFE}
  \begin{split}
    \dot{\gamma}_{a b} \pi^{a b} + \gamma_{a b} \dot{\pi}^{a b} + \frac{\sqrt{\gamma}}{2} \left< \pi \right> \gamma_{a b} \dot{\gamma}^{a b} &\\
    - \sqrt{\gamma} \left< \dot{\gamma}_{a b} \pi^{a b} + \gamma_{a b} \dot{\pi}^{a b} \right> - \frac{\sqrt{\gamma}}{2} \left< \pi \right>  \left< \gamma_{a b} \dot{\gamma}^{a b} \right> &= 0 \,.    
  \end{split}
\end{equation}
Replacing Eqs.\ \eqref{eq:evolg} and \eqref{evolp} in Eq.\ \eqref{eq:LFE} above we get
\begin{equation}
\begin{split}\label{eq:LFE2}
\left( 8 \Delta -2 R - \langle \pi \rangle ^2 + S \right) N &\\
-6 \, \gamma^{-1} \left( \pi^{ab}- \sfrac{1}{3} \pi \gamma^{ab} \right) \left( \pi_{ab}- \sfrac{1}{3} \pi \gamma_{ab} \right) N &= \left\langle \sqrt \gamma ~ \text{\emph{l.h.s.}} \right\rangle \,.
\end{split}
\end{equation}
The term on the right-hand side is a spatial constant and can be written as an undetermined function of time $\varpi(t) = \left\langle \sqrt \gamma ~ \text{\emph{l.h.s.}} \right\rangle$.

\subsection{Spherical symmetry and perfect fluid conditions}\label{sec:sphsym-perf}

We start by applying condition \ref{cond-sphsym} so that our ansatz for the solution is a spherically symmetric metric on 3-space. Following the notation from Ref.\ \cite{Gomes:2015ila} we define
\begin{align}
  \begin{split}
    \diff s^2 =&\, \gamma_{a b} \diff x^a \diff x^b \\
    =&\, \mu^2 (r,t) \diff r^2 + Y^2 (r,t) \diff \Omega^2 \,,
  \end{split}\\
  N =&\, N(r,t) \,,\\
  N^a =&\, \xi (r,t) n^a \,,
\end{align}
where $n^a \equiv \delta_r^a$ is a unit vector in the radial direction and $\diff \Omega^2 \equiv \diff r^2 + r \sin^2 \theta \ud \theta^2$ is the usual 2-sphere element. The canonical momentum conjugate to the metric is defined in spherical symmetry as \cite{Gomes:2015ila}
\begin{equation}
  \pi^{a b} = \diag \left\{ \frac{f}{\mu} \sin \theta, \frac{1}{2} s \sin \theta, \frac{1}{2} \frac{s}{\sin \theta} \right\}
\end{equation}
with $s = s(r,t)$ and $f = f(t,r)$. Spherical symmetry also means that the source matter satisfies the properties
\begin{align}
  \rho =&\, \rho (r,t) \,,\\
  j^a =&\, j (r,t) n^a \,,\\
  S_{a b} =&\, p (r,t) \gamma_{a b} + \sigma (r,t) P_{a b} \,,
\end{align}
where $P_{a b} \equiv \gamma_{a b} - \tensor{\gamma}{_c ^c} n_a n_b$ is the traceless projector with respect to the radial direction. 

Condition \ref{cond-perf} that the source corresponds to a perfect fluid implies that there is no anisotropic stress, that is, in the fluid's rest frame we have
\begin{equation}
  \sigma = 0 \,.
\end{equation}
In addition, condition \ref{cond-perf} limits our choice of fluid, in the sense that it constrains us to place the fluid at rest with respect to the observer, that is,
\begin{equation}\label{eq:comoving}
  j = 0 \,.
\end{equation}
It is worth noting that, while in GR a perfect fluid can always be made comoving with a suitable choice of 4-dimensional coordinates, in SD we have only 3-dimensional diffeomorphism covariance; therefore requiring that the fluid be comoving corresponds to a physical constraint on the matter source.

\section{Solving the constraints}

After applying the conditions from Sec.\ \ref{sec:sphsym-perf} to the constraint equations \eqref{eq:ham-const}, \eqref{eq:mom-const} and \eqref{eq:cmc-cond} we find
\begin{align}
  \begin{split}
     \rho =&\, \frac{1}{\mu^2} \left[ \frac{2}{Y} \left( \frac{\mu'}{\mu} Y' - Y'' \right) - \left( \frac{Y'}{Y} \right)^2 \right] \\
     & - \frac{1}{Y^2} + \frac{f}{2 Y^2} \left( \frac{f}{Y^2} - \frac{s}{\mu} \right) \,, \label{ham-const}
  \end{split}\\
  \frac{f'}{Y} \mu =&\, s Y' \,, \label{mom-const}\\
  \left< \pi \right> =&\, \frac{f}{Y^2} + \frac{s}{\mu} \,, \label{cmc-cond}
\end{align}
and we can immediately solve the CMC condition \eqref{cmc-cond} and momentum constraint \eqref{mom-const} to find
\begin{align}
  s =&\, \mu \left( \left< \pi \right> - \frac{f}{Y^2} \right) \,, \label{CMCs}\\
  f =&\, \frac{\left< \pi \right> Y^2}{3} + \frac{A}{Y} \,, \label{MCf}
\end{align}
where $A (t)$ is a spatially homogeneous integration constant. The homogeneity of $A$ is a consequence of the assumption $j = 0$ in Eq.\ \eqref{eq:comoving} due to the fact that $j$ acts as a source for the radial derivative of $A$. Inserting these results back into the Hamiltonian constraint \eqref{ham-const} we find
\begin{equation}\label{ham-const-cmc}
  \begin{split}
     \rho =&\, \frac{1}{\mu^2} \left[ \frac{2}{Y} \left( \frac{\mu'}{\mu} Y' - Y'' \right) - \left( \frac{Y'}{Y} \right)^2 \right] \\
     & - \frac{1}{Y^2} - \frac{\left< \pi \right>^2}{12} + \frac{3}{4} \frac{A^2}{Y^6} \,.
  \end{split}
\end{equation}

\subsection{Hawking--Hayward mass}\label{sec:ms-mass}

The Hawking--Hayward quasilocal mass \cite{hawking-1968,*Hayward:1993ph} is a measure of the energy content inside a codimension-2 compact hypersurface in general relativity, defined as the Hamiltonian in a $2+2$ foliation of spacetime. It coincides with the Misner-Sharp mass \cite{Misner:1964je} in spherical symmetry and with the Bondi and ADM masses \cite{bondi-1962,*sachs-1962,Arnowitt:1959ah} if the metric is asymptotically flat. Despite the fact that in Shape Dynamics there is no spacetime, it is still useful to use the Hawking--Hayward quasilocal mass as a guide to the determination of the energy content inside a spatial volume. In the following section we use the ADM formalism to cast the Hawking--Hayward mass in terms of quantities contained in the hypersurface, which will render it a meaningful quantity to use in Shape Dynamics. We also interpret the different contributions to the mass in terms of hypersurface quantities. 

It has been pointed out in the literature that in general relativity, since the Hawking--Hayward mass may ultimately be written as a projection of the Riemann tensor, it may therefore be split in two distinct contributions: one from the Ricci tensor and one from the Weyl tensor \cite{Carrera:2009ve,carrera-rmp-2010,Faraoni:2015cnd}. In light of the Einstein equations, these contributions can then be traced as coming from the energy-momentum tensor distributed in the medium (the Ricci part) and from a pointlike source or otherwise compact object (the Weyl part). In Shape Dynamics there is no well-defined spacetime, but once we are in a foliation we may use the ADM splitting to define an analogue to the Hawking--Hayward mass using only quantities defined in the hypersurface. After performing the splitting we may define the Hawking--Hayward mass $M_{\text{HH}}$ as
\begin{equation}\label{ms-ricci-weyl}
  M_\text{HH} = M_\mathcal{R} + M_W \,,
\end{equation}
where the Ricci component $M_\mathcal{R}$ and Weyl component $M_W$, in our spherically symmetric metric, read
\begin{align}
  \begin{split} \label{eq:msr-sphsym}
    M_{\mathcal{R}} =&\, \frac{1}{6 N} \left\{ \xi f' Y - \frac{Y^3}{2 \mu} \left( s \xi \right)' + \frac{Y^4}{\mu} \left( \frac{N'}{\mu Y} \right)' \right.\\
    &\left. + \frac{Y}{\mu} \!\left[ \frac{1}{2} \!\left( Y^2 \dot{s} + N s f \right)\! - \!\dot{f} \mu + N Y \!\left( \frac{Y'}{\mu} \right)' \right] \right\} \\
    & + \frac{1}{3} Y \left( 1 - \frac{Y^{\prime 2}}{\mu^2} \right) - \frac{1}{8} \frac{f^2}{Y} \,,
  \end{split} \\
  \begin{split}\label{eq:msw-sphsym}
    M_{W} =&\, -\frac{1}{6 N} \left\{ \xi f' Y - \frac{Y^3}{2 \mu} \left( s \xi \right)' + \frac{Y^4}{\mu} \left( \frac{N'}{\mu Y} \right)' \right.\\
    &\left. + \frac{Y}{\mu} \!\left[ \frac{1}{2} \!\left( Y^2 \dot{s} + N s f \right)\! - \!\dot{f} \mu + N Y \!\left( \frac{Y'}{\mu} \right)' \right] \right\} \\
    & + \frac{1}{6} Y \left( 1 - \frac{Y^{\prime 2}}{\mu^2} \right) + \frac{1}{4} \frac{f^2}{Y} \,,
  \end{split}
\end{align}
and using Eq.\ \eqref{ms-ricci-weyl} we find that the two parts sum up to
\begin{equation}\label{eq:MSf}
M_\text{HH} = \frac{Y}{2} \left( 1 - \frac{Y^{\prime 2}}{\mu^2} \right) + \frac{f^2}{8 Y} \,,
\end{equation}
as expected \cite{Misra:1973zz}. The full derivation of this result may be found in the Appendix.

\subsection{Solving the Hamiltonian constraint}

An immediate first application of the expression of the Hawking--Hayward mass in spherical symmetry is to algebraically solve the Hamiltonian constraint for $\mu$ without making use of the equations of motion. By substituting the definition from Eq.\ \eqref{eq:MSf} into Eq.\ \eqref{ham-const}, we may cast it as
\begin{equation}
  \rho = \frac{f}{2 Y^2} \left( \frac{s}{\mu} - \frac{f'}{Y Y'} \right) + \frac{2 M'_{\text{HH}}}{Y^2 Y'} \,.
\end{equation}
Applying the momentum constraint \eqref{MCf} and the CMC condition \eqref{CMCs}, we find
\begin{equation}
  \rho = \frac{2 M'_{\text{HH}}}{Y^2 Y'} \,,
\end{equation}
which allows us to write $M_{\text{HH}}$ in integral form in terms of $\rho$ as
\begin{equation}
  M_{\text{HH}} = \eta\,,
\end{equation}
where
\begin{equation}\label{eq:etaprime}
  \eta' = \frac{1}{2} \rho Y^2 Y' \,.
\end{equation}

Inserting this result back into the definition of the Hawking--Hayward mass \eqref{eq:MSf} and noting that it can be solved algebraically for $\mu$, we find
\begin{equation}\label{muHC}
  \mu = \frac{6 Y Y'}{\sqrt{Y^4 \left( \left< \pi \right> + \frac{3 A}{Y^3} \right)^2 + 9 Y \left( 4 Y - 8 \eta \right)}} \,.
\end{equation}
The usefulness of this result will become clear when solving the lapse-fixing equation in Sec.\ \ref{sec:lapsefix}.

\section{Solving the equations of motion}\label{sec:evolution}

The evolution equations for the metric \eqref{eq:evolg} read
\begin{align}
  \dot{Y} =&\, -\frac{N f}{2 Y} + Y' \xi \,, \label{Yponto}\\
  \dot{\mu} =&\, \frac{N}{2} \left( \frac{\mu f}{Y^2} - s \right) + \left( \xi \mu \right)' \,.
\end{align}
Inserting the solutions of the constraint equations, \eqref{CMCs} and \eqref{MCf}, we then find
\begin{align}
  \dot{Y} =&\, - N \left( \frac{\left< \pi \right>}{6} Y + \frac{1}{2} \frac{A}{Y^2} \right) + Y' \xi \,, \\
  \dot{\mu} =&\, - N \left( \frac{\left< \pi \right>}{6} - \frac{A}{Y^3} \right) + \left( \xi \mu \right)' \,.
\end{align}

The presence of a nonzero $A$ implies that there is a singularity in the metric at areal radius $Y = 0$, that is, at both antipodes of the compact spacetime \cite{Gomes:2015ila}. Although condition \ref{sing-center} implies that we have a central singularity, condition \ref{cond-flrw} means that far from the singularity the spacetime is expected to satisfy, among other properties, regularity. Therefore, since $A$ is constant at every leaf due to our choice of $j$ in Eq.\ \eqref{eq:comoving}, in order to guarantee a regular spacetime on the antipodal point we choose $A = 0$. With this choice, the equations of motion simplify significantly, and we now have
\begin{align}
  \frac{\dot{Y}}{Y} =&\, -\frac{\left< \pi \right>}{6} N + \frac{Y'}{Y} \xi \,, \\
  \dot{\mu} =&\, - \frac{\left< \pi \right>}{6} N + \left( \xi \mu \right)' \,, \label{evol-nu-sf}
\end{align}
which also implies that
\begin{equation}\label{metric-evol}
  \frac{\dot{Y}}{Y} - \frac{\dot{\mu}}{\mu} = \xi \left( \frac{Y'}{Y} - \frac{\mu'}{\mu} \right) - \xi' \,.
\end{equation}

If we choose a gauge in which there is no shift (by setting $\xi = 0$), the metric evolution Eq.\ \eqref{metric-evol} results in
\begin{equation}\label{sfY}
  Y = \mu R(r) \,,
\end{equation}
and inserting this result back into Eq.\ \eqref{evol-nu-sf} we find an expression for the lapse, namely
\begin{equation}\label{sfN}
  N = - \frac{6}{\left< \pi \right>} \frac{\dot{\mu}}{\mu} \,.
\end{equation}

We now move on to the momentum equations of motion. The independent components of Eq.\ \eqref{evolp} are
\begin{align}
  \begin{split}
    \dot{f} =&\, N \left(1 + Y^2 p - \frac{3}{4} \frac{f^2}{Y^2} \right) + \xi \left( f' - f \frac{\mu'}{\mu} \right) \\
    &+ f \left( \frac{\dot{\mu}}{\mu} + \frac{s N}{2 \mu} - \xi' \right) - \frac{2 Y Y' N' + N Y^{\prime 2}}{\mu^2}
  \end{split}\\
  \begin{split}
    \dot{s} =&\, N \left[ \mu \left( 2 p + \frac{f^2}{2 Y^4} \right) - \frac{2}{Y} \left( \frac{Y'}{\mu} \right)' \right] + \left( s \xi \right)' \\
    &+ \frac{2 N'}{\mu} \left( \frac{\mu'}{\mu} - \frac{Y'}{Y} \right) - \frac{2 N''}{\mu}    
  \end{split}
\end{align}
Inserting the results from the constraints, Eqs.\ \eqref{MCf} and \eqref{CMCs}, and the equations of motion of the metric, Eqs.\ \eqref{sfY} and \eqref{sfN}, the radial and angular components of the momentum evolution equations read, respectively,
\begin{align}
  \begin{split}
    \frac{1}{3} \dot{\left< \pi \right>} =&\, \frac{6}{\mu^3 \left< \pi \right>} \left[ \frac{\mu'}{\mu} \left( 2 \dot{\mu}' - \frac{\mu' \dot{\mu}}{\mu} \right) + 2 \frac{R'}{R} \dot{\mu}' \right. \\
    & \left. + \frac{\dot{\mu}}{R^2} \left( 1 - R^{\prime 2} \right) \right] + \frac{\dot{\mu}}{\mu} \left( \frac{6 p}{\left< \pi \right>} - \frac{1}{2} \left< \pi \right> \right) \,,
  \end{split}\\
  \begin{split}
    \frac{1}{3} \dot{\left< \pi \right>} =&\, \frac{6}{\mu^3 \left< \pi \right>} \left[ - \frac{\mu'}{\mu} \left( 2 \dot{\mu}' - \frac{\mu' \dot{\mu}}{\mu} \right) + \dot{\mu}'' \right. \\
    & \left. + \frac{1}{R} \left( R' \dot{\mu}' + R'' \dot{\mu} \right) \right] + \frac{\dot{\mu}}{\mu} \left( \frac{6 p}{\left< \pi \right>} - \frac{1}{2} \left< \pi \right> \right) \,.
  \end{split}
\end{align}
The difference between these two components, also known as pressure isotropy condition \cite{stephani-exact}, or spatial Ricci isotropy \cite{carrera-rmp-2010}, yields
\begin{equation}\label{isotropy}
  \begin{split}
    \frac{6}{\mu^3 \left< \pi \right>} \left[ \dot{\mu}'' + 2 \frac{\mu'}{\mu} \left( 2 \dot{\mu}' - \frac{\mu' \dot{\mu}}{\mu} \right) - \frac{R'}{R} \dot{\mu}' \right.& \\
    \left. + \frac{1}{R^2} \left( R R'' - R^{\prime 2} + 1 \right) \right] &= 0 \,,
  \end{split}
\end{equation}
so we are motivated to choose a gauge for which \cite{sussman-1987,mcvittie-1984}
\begin{align}
  R R'' - R^{\prime 2} + 1 =&\, 0 \quad \Rightarrow \quad R^{\prime 2} = 1 - k R^2 \label{Rdef}\\
\intertext{so that}
R =&\,
\begin{cases}
  \sinh r & k = -1 \,, \\
  r & k = 0 \,, \\
  \sin r & k = 1 \,.
\end{cases} \label{Rcases}
\end{align}
Our gauge choice may be further specified by choosing a value for $k$, which will remain undetermined in our analysis. For $\dot{\mu} \neq 0$, Eq.\ \eqref{isotropy} can then be rewritten as a total derivative, namely
\begin{equation}
  \partial_t \left( \mu'' -2 \frac{\mu^{\prime 2}}{\mu} - \mu' \frac{R'}{R} \right) = 0 \,,
\end{equation}
which reduces to the ordinary differential equation
\begin{equation}\label{isotropy_R}
  \mu'' -2 \frac{\mu^{\prime 2}}{\mu} - \mu' \frac{R'}{R} = \chi(r) \,.
\end{equation}

Inserting these results back into the Hamiltonian constraint \eqref{ham-const}, we find
\begin{equation}\label{ham-const-sf}
  \begin{split}
    \rho =&\, - \frac{ \left< \pi \right>^2}{12} + \frac{3}{\mu^2} \left\{ \frac{1 - R^{\prime 2}}{R^2} \right. \\
    & \left. - \frac{2}{\mu} \left[ \frac{\chi}{3} + \mu' \left( \frac{R'}{R} + \frac{\mu'}{\mu} \right) \right] \right\} \,,
  \end{split}
\end{equation}
and taking the radial derivative of Eq.\ \eqref{ham-const-sf} and using again Eqs.\ \eqref{isotropy_R} and \eqref{Rdef}, we find
\begin{equation}\label{eq:chiprime}
  \chi' + 3 \chi \frac{R'}{R} = - \frac{1}{2} \rho' \mu^3 \,,
\end{equation}
whose solution is
\begin{equation}
  \chi = \frac{1}{R^3} \left[ -\frac{1}{2}  \int \rho' \left( \mu R \right)^3 \diff r + m_0 \right] \,.
\end{equation}

We may use the results from this section to rewrite the Hawking--Hayward mass in terms of the new metric functions. After applying the results from the constraint equations, \eqref{CMCs} and \eqref{MCf}, and the equations of motion \eqref{sfY}, \eqref{sfN} and \eqref{isotropy_R}, we find that the two parts of Eq.\ \eqref{ms-ricci-weyl} reduce to
\begin{align}
  \begin{split}
    M_{\mathcal{R}} =&\, \frac{\mu R}{2} \left( 1 - R^{\prime 2} \right) - R^2 \mu' R' \\
    & + R^3 \left( \frac{\left< \pi \right>^2 \mu^3}{72} - \frac{\chi}{3} - \frac{\mu^{\prime 2}}{2 \mu} \right) \,,    
  \end{split}\\
  M_W =&\, \frac{R^3}{3} \chi \,,
\end{align}
with $\chi$ given by Eq.\ \eqref{isotropy_R}. We may interpret this result as the fact that a nonzero $\chi$ corresponds to a contribution to the energy of the system from a source other than the fluid, such as a compact object or a black hole, as we will see in the following section.

Now, inspecting Eq.\ \eqref{eq:chiprime}, and using Eq.\ \eqref{sfY} we notice that $\chi$ and $\eta$ defined in Eq.\ \eqref{eq:etaprime} are related by
\begin{equation}\label{eq:etachi}
  R^3 \chi = 6 \eta - \rho R^3 \mu^3 + B(t) \,,
\end{equation}
where $B(t)$ is an integration constant. Notice that the left-hand side is independent of time.

By inserting the results of the pressure isotropy condition, Eqs.\ \eqref{Rdef} and \eqref{isotropy_R} into one of the momentum equations of motion, we find 
\begin{equation}
  \begin{split}
    p =&\, -\frac{1}{12} \left< \pi \right>^2 - \frac{1}{18} \frac{\mu}{\dot{\mu}} \dot{\left< \pi \right>} \left< \pi \right> + 2 \frac{\dot{\mu}'}{\mu^2 \dot{\mu}} \left( \frac{R'}{R} + \frac{\mu'}{\mu} \right) \\
    &- \frac{1}{\mu^2} \left( k + \frac{\mu^{\prime 2}}{\mu^2} \right) \,,
  \end{split}
\end{equation}
which may interpret as the definition of the fluid pressure $p$.

\section{Finding a particular solution}

The Ricci part $M_{\mathcal{R}}$ of the Hawking--Hayward quasilocal mass is associated with the presence of a source field in the equations of motion, since it vanishes in a vacuum solution. The Weyl part $M_W$ is therefore associated with the presence of a central compact object, as well as the presence of spacetime singularities. In particular, we might choose a finite interval for the radial coordinate by setting $k = 1$ in Eq.\ \eqref{Rcases}, which, for particular forms of $\mu$ would be akin to closed FLRW models. In this case, for the foliation to be regular at the antipode $r = \pi$, in addition to setting $A = 0$ on Sec.\ \ref{sec:evolution}, the Weyl tensor must also be finite at the antipode. On the other hand, at the coordinate center $r = 0$ a singularity is guaranteed if the Weyl tensor diverges, therefore satisfying condition \ref{sing-center}. The simplest choice of a function $\chi$ which satisfies both requirements while relying only on the already defined gauge-fixing function $R$ is given by
\begin{equation}\label{chi-MS}
  \chi = 3 m \frac{w^{\prime 2}}{w^3} \,,
\end{equation}
where $m$ is a constant, and
\begin{equation}
  w(r) \equiv 2 R \left( \frac{r}{2} \right) \,.
\end{equation}
By making this choice we have both fixed the mass of the central object $M_W$ to be equal to $m$\footnote{This interpretation becomes clear if one chooses a spatially flat asymptotic spacetime by setting $k = 0$ in Eq.\ \eqref{Rcases}.}, as well as ensured that condition \ref{sing-center} holds.

Applying Eq.\ \eqref{chi-MS} to Eq.\ \eqref{isotropy_R}, and noting that Eq.\ \eqref{Rdef} implies that $w^{\prime 2} = 1 - \frac{k}{4} w^2$, we have
\begin{equation}\label{isotropy-w}
  \mu'' -2 \frac{\mu^{\prime 2}}{\mu} - \mu' \frac{R'}{R} = 3 m \frac{w^{\prime 2}}{w^3} \,.
\end{equation}
The pressure isotropy condition \eqref{isotropy-w} now admits a solution of the form
\begin{equation}
  \mu = a \left( 1 + \frac{m}{2 a w} \right)^2 \,,
\end{equation}
with $a = a(t)$ an arbitrary function of time only. Now that we have both $\chi$ and $\mu$, we can calculate the Hawking--Hayward mass of this solution, and we get
\begin{equation}
  M_\text{HH} = \frac{w^3}{8} \left( 4 - k w^2 \right)^{\frac{3}{2}} \left[ \frac{a k}{2} + \frac{H^2 a^3}{2} \left( 1 + \frac{m}{2 a w} \right)^6 + \frac{m}{w^3} \right] \,,
\end{equation}
and Eq.\ \eqref{eq:etachi} results in $B = 0$.

Eq.\ \eqref{sfN} now yields
\begin{equation}\label{eq:LFE-mcvittie-sol}
  N = -\frac{6}{\left< \pi \right>} \frac{1 - \frac{m}{2 a w}}{1 + \frac{m}{2 a w}} \frac{\dot{a}}{a} \,,
\end{equation}
and for condition \ref{cond-flrw} to hold the solution requires that $N \to 1$ as $r \gg m$, and since this must be satisfied for all slices we are required to choose $\left<\pi \right> = -6 \frac{\dot{a}}{a}$. The 4D metric finally reads
\begin{subequations}\label{mcvittie-sol}
\begin{align}
  \diff s^2 =&\, a(t)^2 \left[ 1 + \frac{m}{2 a (t) \, w(r)} \right]^4 \! \left[ \diff r^2 + R (r)^2 \diff \Omega^2 \right],\\
  N =&\, \frac{1 - \frac{m}{2 a(t) \, w(r)}}{1 + \frac{m}{2 a(t) \, w(r)}} \,,\\
  N^i =&\, 0 \,.
\end{align}
\end{subequations}
This is the well-known McVittie metric \cite{McVittie:1933zz,*mcvittie-1932} written in a compact foliation.

\subsection{Lapse-fixing equation}\label{sec:lapsefix}

Rewriting Eq.~\eqref{eq:LFE2} in our spherically symmetric ansatz and after applying the solutions of the CMC \eqref{CMCs}, Hamiltonian \eqref{ham-const} and momentum \eqref{MCf} constraints,
\begin{equation}
  \begin{split}
    -\frac{\varpi}{4} =&\, N \left( \rho + 3 p + \frac{\left< \pi \right>}{6} + \frac{2 A^2}{Y^6} \right) \\
    & + \frac{2 N'}{\mu^2} \left( \frac{\mu'}{\mu} - \frac{2 Y'}{Y} \right) - \frac{2 N''}{\mu^2} \,.
  \end{split}
\end{equation}
Using Eq.\ \eqref{muHC} we may  write it as
\begin{equation}\label{eq:lfe}
  \begin{split}
    \frac{1}{2 Y^2 Y'} \left[ \frac{4 Y^2 Y^{\prime 2}}{\mu^3} \left( \frac{N \mu}{Y'} \right)' \right]'&\\
    - N \left[ 3 \left( \rho + p \right) + \frac{Y \rho'}{Y'} \right] &= \frac{\varpi}{4} \,.
  \end{split}
\end{equation}
After applying the results from gauge-fixing the metric evolution equations [Eqs.\ \eqref{sfY} and \eqref{sfN}] and the pressure isotropy condition [Eqs.\ \eqref{Rcases} and \eqref{isotropy_R}] we find
\begin{equation}
  \begin{split}
    \frac{\varpi \left< \pi \right>}{24} \frac{\mu}{\dot{\mu}} =&\, \frac{1}{R' \mu + R \mu'} \left( \frac{6 \chi R'+ 2 R \chi'}{\mu^2} + R \rho' \mu \right) \\
    & + \frac{6}{\mu^2} \left[ \frac{R'}{R} \frac{\mu'}{\mu} + \frac{\chi}{\mu} - k + \frac{2 \mu' - 1}{\mu} \left( \frac{R'}{R} + \frac{\mu'}{\mu} \right) \right] \\
& + 3 \left( \rho + p \right) \,.
  \end{split}
\end{equation}
Inserting the solution \eqref{mcvittie-sol} into the lapse-fixing equation, we find that it is identically satisfied. This proves that the McVittie lapse~\eqref{eq:LFE-mcvittie-sol} is a solution  of the lapse fixing equation~\eqref{eq:LFE2}. Our solution is consistent and is (locally) both a solution of Shape Dynamics and General Relativity.

\section{Conclusions}\label{sec:conclusions}

In this work we have found a new solution of Shape Dynamics as part of potentially an entire new class of solutions to this theory, by building on the fact that in spherical symmetry a CMC foliation is equivalent to a shearfree comoving flow. There are many implications of this result, and the next step is now to fully characterize this solution. Although we know the causal structure of spatially flat McVittie spacetimes in general relativity \cite{Kaloper:2010ec,*Lake:2011ni,daSilva:2012nh}, the spatially compact counterpart, despite having been previously studied in the context of general relativity \cite{sussman-1988b,*Nolan:1998xs}, requires a completely new interpretation in the context of Shape Dynamics.

We still do not know whether the class of metrics we studied in fact contains a black hole. The 4-dimensional compact McVittie metric does not possess an event horizon, but it may contain apparent horizons which are not covered by the coordinate patch used in previous analyses \cite{sussman-1988b}. As in previously studied Shape Dynamics solutions, the event horizon or trapping surface may well give way to a throat into another region of space which possesses no general-relativistic analogue \cite{Gomes:2015ila}, which may extend the explorable region until a possible universal horizon or otherwise locus where the CMC foliation can no longer be extended \cite{Maciel:2015ypv,*Meiers:2015rzm}. Also, regarding singularities, the known McVittie spacetime singularities all stem from divergences of 4-dimensional quantities, in regions where often 3-dimensional quantities remain well-behaved. Therefore, they may very well represent perfectly regular and traversable regions in a Shape Dynamics interpretation.

Another important question that has been left open here is the action of the matter source. It may be possible to follow a similar procedure to previous Hamiltonian analyses of $k$-essence models \cite{Abramo:2005be} in order to find a generic perfect-fluid source, so we may characterize the source from a field theory perspective and provide an analogy with the cuscuton source of the McVittie spacetime \cite{Abdalla:2013ara}.

Finally, it must be noted that a much broader class of solutions of Eq.\ \eqref{isotropy_R} has been studied in Ref.\ \cite{wyman-1976}. Their applicability as solutions of Shape Dynamics, as well as whether they can be related to general solutions of the lapse-fixing equation \eqref{eq:lfe}, will be the object of future work.

\begin{acknowledgments}

We thank N.\ Afshordi, H.\ A.\ Gomes, S.\ Gryb, A.\ Maciel, L.\ Smolin and R.\ Sorkin for insights and valuable discussions. DCG is supported by CNPq Grant No.\ 206101/2014-7. This research was supported in part by Perimeter Institute for Theoretical Physics. Research at Perimeter Institute is supported by the Government of Canada through Industry Canada and by the Province of Ontario through the Ministry of Economic Development \& Innovation. This research was also partly supported by a grant from the John Templeton Foundation. 
  
\end{acknowledgments}

\appendix*

\section{Interpretation of the Hawking mass as a two-component quantity}\label{sec:appendix-ms}

In this Appendix we perform the derivation and split of the Hawking--Hayward mass from 4-dimensional spacetime in terms of 3-dimensional quantities defined in our spacelike foliation. To do so we use some of the techniques developed in Refs.\ \cite{Faraoni:2015cnd,carrera-rmp-2010,*Carrera:2009ve}. To avoid cluttering the notation, all quantities represent 4-dimensional objects unless stated otherwise. Greek indices run from 0 to 3 and we use the signature $(-,+,+,+)$.

We start by defining a few projectors into the hypersurface. The comoving flow of a 4-dimensional metric $g$ is defined as
\begin{equation}\label{flow-def}
    u^\mu \equiv \frac{1}{\sqrt{-g_{00}}} \delta_t^\mu \,,
\end{equation}
and the orthogonal projection with respect to $u^\mu$ gives the spatial slices
\begin{equation}\label{3metric-def}
  \gamma_{\mu \nu} \equiv g_{\mu \nu} + u_\mu u_\nu \,.
\end{equation}
We may also define a unit vector $n^\mu$ orthogonal to the flow. To do so, we use the acceleration $\dot{u}_\mu \equiv u^\alpha u_{\mu; \alpha}$ to write $n^\mu \equiv \frac{\dot{u}^{\mu}}{\sqrt{\dot{u}^\alpha \dot{u}_\alpha}}$, which we use to construct the induced metric on a codimension-2 hypersurface,
\begin{equation}\label{2metric}
  h_{\mu \nu} \equiv \gamma_{\mu \nu} - n_\mu n_\nu \,,
\end{equation}
an example of which in spherical symmetry is a 2-sphere of radius $r$.

\subsection{Hypersurface decomposition}\label{sec:ms-hyper}

The Hawking--Hayward quasilocal mass is defined in terms of radial null geodesics on a compact spacelike 2-surface $\mathcal{S}$ as the integral of the Hamiltonian two-form over the surface $\mathcal{S}$, multiplied by the length scale given by the area of $\mathcal{S}$ \cite{hawking-1968,*Hayward:1993ph}
\begin{equation}\label{mhh2null}
  \begin{split}
    M_\text{HH} \equiv&\, \frac{1}{(4 \pi)^{\nicefrac{3}{2}}} \frac{\sqrt{A}}{4} \int_\mathcal{S} \ud \mathcal{S} \left[ \vphantom{\frac{1}{2}} \tensor[^{(2)}]{\mathcal{R}}{} + \theta_{(+)} \theta_{(-)} \right. \\
    &\left. - \frac{1}{2} \sigma^{(+)}_{\mu \nu} \sigma^{\mu \nu}_{(-)} - 2 \omega_\mu \omega^\mu \right] \,,
  \end{split}
\end{equation}
where $A \equiv \int_\mathcal{S} \ud \mathcal{S}$ is the area of $\mathcal{S}$, $\theta_{(\pm)}$ are the expansion scalars of the ingoing ($-$) and outgoing ($+$) null geodesics, $\sigma^{(\pm)}_{\mu \nu}$ is their respective shear tensor of the geodesic congruence, and $\omega^{\mu}$ the twist of the surface $\mathcal{S}$. In spherical symmetry $\omega^{\mu}$ vanishes, and we may use the contracted Gauss equation to write
\begin{equation}\label{cont-gauss}
  \tensor[^{(2)}]{\mathcal{R}}{} + \theta_+ \theta_- - \frac{1}{2} \sigma^{(+)}_{\mu \nu} \sigma^{\mu \nu}_{(-)} = h^{\alpha \gamma} h^{\beta \delta} \mathcal{R}_{\alpha \beta \gamma \delta} \,,
\end{equation}
with $h_{\mu \nu}$ as defined in Eq.\ \eqref{2metric}. In this context, we can use Eq.\ \eqref{cont-gauss} to rewrite the Hawking--Hayward mass from Eq.\ \eqref{mhh2null} as
\begin{equation}\label{eq:mhh4riem}
  M_\text{HH} = \frac{1}{(4 \pi)^{\nicefrac{3}{2}}} \frac{\sqrt{A}}{4} \int_\mathcal{S} \diff \mathcal{S} \, h^{a c} h^{b d} \mathcal{R}_{a b c d} \,.
\end{equation}

Moreover, using the decomposition of the Riemann tensor into its Ricci and Weyl parts, that is,
\begin{equation}\label{eq:riem-ricci-weyl}
  \begin{split}
    \mathcal{R}_{\mu \nu \alpha \beta} =&\, C_{\mu \nu \alpha \beta} + g_{\mu \left[ \alpha \right.} \mathcal{R}_{\left. \beta \right] \nu} - g_{\nu \left[ \alpha \right.} \mathcal{R}_{\left. \beta \right] \mu} \\
    & - \frac{\mathcal{R}}{3} g_{\mu \left[ \alpha \right.} g_{\left. \beta \right] \nu} \,,
  \end{split}
\end{equation}
we finally recover Eq.\ \eqref{ms-ricci-weyl} with the contributions from the Ricci part and Weyl part defined as
\begin{align}\label{MHHricci}
  \begin{split}
    M_\mathcal{R} \equiv&\, \frac{1}{(4 \pi)^{\nicefrac{3}{2}}} \frac{\sqrt{A}}{4} \int_\mathcal{S} \ud \mathcal{S} h^{\mu \alpha} h^{\nu \beta} \left( g_{\mu \left[ \alpha \right.} \mathcal{R}_{\left. \beta \right] \nu} \right. \\
    & \left. - g_{\nu \left[ \alpha \right.} \mathcal{R}_{\left. \beta \right] \mu} - \frac{\mathcal{R}}{3} g_{\mu \left[ \alpha \right.} g_{\left. \beta \right] \nu} \right) \,,
  \end{split}\\
  \begin{split}\label{MHHweyl}
    M_W \equiv&\, \frac{1}{(4 \pi)^{\nicefrac{3}{2}}} \frac{\sqrt{A}}{4} \int_\mathcal{S} \ud \mathcal{S} h^{\mu \alpha} h^{\nu \beta} C_{\mu \nu \alpha \beta} \,.
  \end{split}
\end{align}
In fact, the integrand in Eq.\ \eqref{MHHweyl} is the electric part of the Weyl tensor, which prompts the interpretation of a ``Newtonian'' character of the Hawking--Hayward mass \cite{Faraoni:2015cnd}.

\subsection{Back to the hypersurface}

Using the contracted Gauss equation, the Hawking--Hayward quasilocal mass can be cast entirely in terms of quantities within a 3-dimensional surface \cite{Faraoni:2015cnd}. Since the mass depends on the projection of the Riemann tensor on the 2-surface, it is useful to compute the projection of Eq.\ \eqref{eq:riem-ricci-weyl} on the 2-surface, that is,
\begin{equation}
  \begin{split}
   h^{a b} h^{c d} \mathcal{R}_{a b c d} = &\, h^{a b} h^{c d} C_{a b c d} - \frac{\mathcal{R}}{3} h^{a b} h^{c d} g_{a \left[ c \right.} g_{\left. d \right] b} \\
    & + h^{a b} h^{c d} \left( g_{a \left[ c \right.} \mathcal{R}_{\left. d \right] b} - g_{b \left[ c \right.} \mathcal{R}_{\left. d \right] a} \right) \,.
  \end{split}
\end{equation}
We start by noticing the following identity:
\begin{align}
  h^{a b} h^{c d} g_{a \left[ c \right.} g_{\left. d \right] b} =&\, 1 \,.
\end{align}
Additionally, for any symmetric rank-2 tensor $T$, we have
\begin{equation}
  h^{a b} h^{c d} \left( g_{a \left[ c \right.} T_{\left. d \right] b} - g_{b \left[ c \right.} T_{\left. d \right] a} \right) = h^{a b} T_{a b} \,,
\end{equation}
Making use of these results, the 2-surface projected Riemann tensor reads
\begin{equation}
  h^{a b} h^{c d} \mathcal{R}_{a b c d} = h^{a b} h^{c d} C_{a b c d} + h^{a b} \mathcal{R}_{a b} - \frac{\mathcal{R}}{3} \,.
\end{equation}
Thus, we may cast the individual contributions from Sec.\ \ref{sec:ms-hyper} by writing the contributions from Eqs.\ \eqref{MHHricci} and \eqref{MHHweyl} as
\begin{align}
  \begin{split}\label{eq:msr-4r}
    M_{\mathcal{R}} =&\, \frac{1}{(4 \pi)^{\nicefrac{3}{2}}} \frac{\sqrt{A}}{4} \int_\mathcal{S} \diff \mathcal{S} \, \left( h^{a b} \mathcal{R}_{a b} - \frac{\mathcal{R}}{3} \right) \,,
  \end{split}\\
  \begin{split}\label{eq:msw-4r}
    M_{W} =&\, \frac{1}{(4 \pi)^{\nicefrac{3}{2}}} \frac{\sqrt{A}}{4} \int_\mathcal{S} \diff \mathcal{S} \, \left( h^{a b} h^{c d} \mathcal{R}_{a b c d} \vphantom{\frac{\mathcal{R}}{3}} \right.\\
    &\left. - h^{a b} \mathcal{R}_{a b} + \frac{\mathcal{R}}{3} \right) \,.
  \end{split}
\end{align}

Using the contracted Gauss-Codazzi equations, we write the Ricci scalar and projected Ricci tensor in ADM form as
\begin{align}
  \begin{split}\label{eq:projriccit}
    h^{a b} \tensor{\mathcal{R}}{_{a b}} =&\, h^{a b} \left[ \tensor[^{(3)}]{\mathcal{R}}{_{a b}} + K K_{a b} - 2 \tensor{K}{^c _b} K_{c a} \right.\\
    &\left. - \frac{1}{N} \left( \partial_t K_{a b} - \mathcal{L}_{\xi} K_{a b} + \nabla_a \nabla_b N \right) \right] \,,     
  \end{split}\\
  \begin{split}\label{eq:projriccisc}
    \tensor{\mathcal{R}}{} =&\, \tensor[^{(3)}]{\mathcal{R}}{} + K^2 - 3 K^{a b} K_{a b} \\
    & - \frac{2}{N} \gamma^{a b} \left( \partial_t K_{a b} - \mathcal{L}_{\xi} K_{a b} + \nabla_a \nabla_b N \right) \,,
  \end{split}
\end{align}
and we use Eqs.\ \eqref{eq:projriccit} and \eqref{eq:projriccisc} to cast the Ricci and Weyl contributions from Eqs.\ \eqref{eq:msr-4r} and \eqref{eq:msr-4r} as
\begin{widetext}
\begin{align}
  \begin{split}
    M_{\mathcal{R}} =&\, \frac{1}{(4 \pi)^{\nicefrac{3}{2}}} \frac{\sqrt{A}}{4} \int_\mathcal{S} \diff \mathcal{S} \, \left\{ h^{a b} \left[ \tensor[^{(3)}]{\mathcal{R}}{_{a b}} + K K_{a b} - 2 \tensor{K}{^c _b} K_{c a} - \frac{1}{N} \left( \dot{K}_{a b} - \mathcal{L}_{\xi} K_{a b} + \nabla_a \nabla_b N \right) \right] \right. \\
    & \left. - \frac{1}{3} \left[ \tensor[^{(3)}]{\mathcal{R}}{} + K^2 - 3 K^{a b} K_{a b} - \frac{2}{N} \gamma^{a b} \left( \dot{K}_{a b} - \mathcal{L}_{\xi} K_{a b} + \nabla_a \nabla_b N \right) \right] \right\} \,,
  \end{split}\\
  \begin{split}
    M_{W} =&\, \frac{1}{(4 \pi)^{\nicefrac{3}{2}}} \frac{\sqrt{A}}{4} \int_\mathcal{S} \diff \mathcal{S} \, \left\{ h^{a c} h^{b d} \left( \tensor[^{(3)}]{\mathcal{R}}{_{a b c d}} + K_{a c} K_{b d} - K_{a d} K_{b c} \right) \vphantom{\frac{1}{N}} \right.\\
    & - h^{a b} \left[ \tensor[^{(3)}]{\mathcal{R}}{_{a b}} + K K_{a b} - 2 \tensor{K}{^c _b} K_{c a} - \frac{1}{N} \left( \dot{K}_{a b} - \mathcal{L}_{\xi} K_{a b} + \nabla_a \nabla_b N \right) \right] \\
    & \left. + \frac{1}{3} \left[ \tensor[^{(3)}]{\mathcal{R}}{} + K^2 - 3 K^{a b} K_{a b} - \frac{2}{N} \gamma^{a b} \left( \dot{K}_{a b} - \mathcal{L}_{\xi} K_{a b} + \nabla_a \nabla_b N \right) \right] \right\} \,.
  \end{split}
\end{align}
\end{widetext}
which depend solely on quantities defined in the hypersurface. By applying this result to our spherically symmetric ansatz we arrive at the expressions from Eqs.\ \eqref{eq:msr-sphsym} and \eqref{eq:msw-sphsym}.

\bibliography{shortnames,referencias}

\end{document}